\documentclass[11pt]{article}
\usepackage{amsmath,amssymb,amsfonts,bbm}
\usepackage{multirow}
\usepackage[table,xcdraw]{xcolor}
\usepackage{graphicx}
\usepackage{caption}
\usepackage{subcaption}
\usepackage[export]{adjustbox}
\usepackage{setspace}
\usepackage{float}
\DeclareGraphicsExtensions{.pdf,.png,.jpg}
\DeclareCaptionFormat{subfig}{\figurename~#1#2#3}
\DeclareCaptionSubType*{figure}
\newcommand{\be}{\begin{equation}}
\newcommand{\ee}{\end{equation}}
\newcommand{\bea}{\begin{eqnarray}}
\newcommand{\eea}{\end{eqnarray}}
\newcommand{\ba}{\begin{array}}
\newcommand{\ea}{\end{array}}

\newcommand{\htwo}{h_{2,1}}
\newcommand{\E}{\mathcal{E}}
\newcommand{\M}{\mathcal{M}}
\newcommand{\N}{\mathcal{N}}

\newcommand{\K}{\mathcal{K}}

\long\def\symbolfootnote[#1]#2{\begingroup%
\def\thefootnote{\fnsymbol{footnote}}\footnote[#1]{#2}\endgroup}

\begin{document}

\thispagestyle{empty}\vspace{40pt}

\hfill{}

\vspace{128pt}

\begin{center}
    \textbf{\Large The geodesic structure of BPS one-branes in five dimensions}\\
    \vspace{40pt}

      Tahia F. Dabash$^{a, b}$\symbolfootnote[2]{\tt tahia.dabash@science.tanta.edu.eg}, Moataz H. Emam$^{c}$\symbolfootnote[3]{\tt moataz.emam@cortland.edu}

\end{center}

    \vspace{3pt}
    \begin{changemargin}{0.5in}{0in}
    \begin{flushleft}
        \begin{spacing}{1.0}
    $^a$ \textit{\small  Department of Mathematics, Tanta University, Tanta, Egypt}\\
    $^b$ \textit{\small  Egyptian Relativity Group (ERG)}\\
    $^c$ \textit{\small  Department of Physics, SUNY Cortland, Cortland, New York, 13045, USA}\\
        \end{spacing}
    \end{flushleft}
    \end{changemargin}

\vspace{6pt}

\begin{abstract}
    In this paper, we continue previous work where one-brane spacetimes coupled to the $\N=2$ ungauged five dimensional hypermultiplets were found. We explore their symmetries as well as study their full geodesic structure. The one-branes are characterized by a coupling constant that distinguishes the behavior of the geodesics from smooth and causally connected in the positive case to singular and repulsive in the negative case.
\end{abstract}

\newpage



\vspace{15pt}

\pagebreak

\section{Introduction}

Studies of $\N=2$ supergravity theories in four and five dimensions usually focus on the vector and/or tensor multiplets regimes. This is due to the fact that their underlying special K\"{a}hler geometry is very well understood (see, for example, \cite{Butter:2012xg,Klemm:2009uw,Mohaupt:2008zz,Cacciatori:2008ek,Cortes:2003zd,Isozumi:2003uh,Cacciatori:2003kv,Andrianopoli:2011zj} and references within). On the other hand, solutions in the hypermultiplets sector are rare due in part to the mathematical complexity involved since the hypermultiplets generally parameterize quaternionic manifolds \cite{de Wit:2001dj}. However, it was pointed out some years ago that due to the so-called $c$-map, the hypermultiplets in $D=5$, for instance, can be related to the much better-understood $D=4$ vector multiplets, and that the methods of special geometry developed for the latter, can be applied to the former \cite{Gutperle:2000ve}. Based on this observation, some hypermultiplet constructions in instanton and certain brane backgrounds were found and studied (last reference and \cite{Emam:2005bh,Emam:2006sr}). More recently, one of us argued in \cite{Emam:2009xj} that the well-known symplectic structure of quaternionic and special  K\"{a}hler manifolds \cite{deWit:1995jd} can be used to construct hypermultiplet ``solutions'' based on covariance in symplectic space. These are full solutions \emph{only} in the symplectic sense, written in terms of symplectic basis vectors and invariants. Using these methods, a solution was found in \cite{Emam:2013mq} that represents $D=5$ Bogomol'nyi-Prasad-Sommerfield (BPS) one-branes coupled to the full set of hypermultiplet fields. Explicit spacetime solutions to the hyperscalars were found, as well as constraints on the complex structure moduli of the underlying Calabi-Yau. These spacetimes represent string-like 1-branes in $D=5$ characterized by a coupling constant $q\in \mathbb{R}$. In this paper, we continue the study of these solutions by first explicitly finding and classifying their Killing vector fields, which we find to be the expected translational and rotational ones. We then proceed to calculate and plot the geodesics of these spacetimes and find that they show a smooth behavior at positive $q$ while in the negative case there exists a singular spherical shell at a specific radius from the brane, which acts as a repulsive center to the geodesics. In the literature, similar calculations were performed for a variety of situations, for example \cite{Gonzalez:2023jhx}, \cite{Kubiznak:2011ay}, \cite{Frolov:2003en}, \cite{Hackmann:2008tu}, \cite{Kagramanova:2012hw}, \cite{Gonzalez:2015qxc}, and \cite{Chandler:2015aha}.

\section{One branes in $D=5$ $\N=2$ supergravity with hypermultiplets} \label{theory}

The dimensional reduction of $D=11$ supergravity theory over a Calabi-Yau 3-fold $\M$ with nontrivial complex structure moduli yields an $\N=2$ supergravity theory in $D=5$ with a set of scalar fields and their supersymmetric partners all together known as the \emph{hypermultiplets}. These are partially comprised of the \emph{universal hypermultiplet} $\left(a, \sigma, \zeta^0, \tilde \zeta_0\right)$, so called because it appears irrespective of the detailed structure of $\M$. The field $a$ is known as the universal axion and the dilaton $\sigma$ is proportional to the natural logarithm of the volume of $\M$. The rest of the hypermultiplets are $\left(z^i, z^{\bar i}, \zeta^i, \tilde \zeta_i: i=1,\ldots, \htwo\right)$, where the $z$'s are identified with the complex structure moduli of $\M$, and $\htwo$ is the Hodge number determining the dimensions of the manifold of the Calabi-Yau's complex structure moduli; $\M_C$. The `bar' over an index denotes complex conjugation. The fields $\left(\zeta^I, \tilde\zeta_I: I=0,\ldots,\htwo\right)$ are known as the axions and arise as a result of the $D=11$ Chern-Simons term. The supersymmetric partners known as the hyperini complete the hypermultiplets. The axionic fields $\left(\zeta^I, \tilde\zeta_I\right)$ can be defined as components of a symplectic vector $\left| \Xi  \right\rangle$ such that the symplectic scalar product is defined by, for example:
\be
    \left\langle {{\Xi }}
 \mathrel{\left | {\vphantom {{\Xi } d\Xi }}
 \right. \kern-\nulldelimiterspace}
 {d\Xi } \right\rangle   = \zeta^I d\tilde \zeta_I  - \tilde \zeta_I
 d\zeta^I,\label{DefOfSympScalarProduct}
\ee
where $d$ is the spacetime exterior derivative $\left(d=dx^\mu\partial_\mu:\mu=0,\ldots,4\right)$. One can define the symplectic basis vectors $\left| V \right\rangle $, $\left| {U_i } \right\rangle $ and their complex conjugates such that
\bea
    \left\langle {{\bar V}}
     \mathrel{\left | {\vphantom {{\bar V} V}}
     \right. \kern-\nulldelimiterspace}
     {V} \right\rangle   &=& i, \quad\quad
    \left|\nabla _i  {\bar V} \right\rangle  = \left|\nabla _{\bar i}  V \right\rangle =0\nonumber\\
    \left\langle {{U_i }}
    \mathrel{\left | {\vphantom {{U_i } {U_j }}}
    \right. \kern-\nulldelimiterspace}
    {{U_j }} \right\rangle  &=& \left\langle {{U_{\bar i} }}
    \mathrel{\left | {\vphantom {{U_{\bar i} } {U_{\bar j} }}}
    \right. \kern-\nulldelimiterspace}
    {{U_{\bar j} }} \right\rangle    =
    \left\langle {\bar V}
    \mathrel{\left | {\vphantom {\bar V {U_i }}}
    \right. \kern-\nulldelimiterspace}
    {{U_i }} \right\rangle  = \left\langle {V}
    \mathrel{\left | {\vphantom {V {U_{\bar i} }}}
    \right. \kern-\nulldelimiterspace}
    {{U_{\bar i} }} \right\rangle  = \left\langle { V}
    \mathrel{\left | {\vphantom { V {U_i }}}
    \right. \kern-\nulldelimiterspace}
    {{U_i }} \right\rangle=\left\langle {\bar V}
    \mathrel{\left | {\vphantom {\bar V {U_{\bar i} }}}
    \right. \kern-\nulldelimiterspace}
    {{U_{\bar i} }} \right\rangle= 0,\nonumber\\
    \left|\nabla _{\bar j}  {U_i } \right\rangle  &=& G_{i\bar j} \left| V \right\rangle ,\quad \quad \left|\nabla _i  {U_{\bar j} } \right\rangle  = G_{i\bar j} \left| {\bar V}
    \right\rangle,\quad\quad
    G_{i\bar j}= \left( {\partial _i \partial _{\bar j} \K} \right)=- i    \left\langle {{U_i }}
    \mathrel{\left | {\vphantom {{U_i } {U_{\bar j} }}}
    \right. \kern-\nulldelimiterspace}
    {{U_{\bar j} }} \right\rangle,\nonumber\\
    {\bf \Lambda } &=& 2\left| V \right\rangle \left\langle {\bar V} \right| + 2G^{i\bar j} \left| {U_{\bar j} } \right\rangle \left\langle {U_i } \right|-i
\eea
where the derivatives are with respect to the moduli $\left(z^i, z^{\bar i}\right)$, $G_{i\bar j}$ is a special K\"{a}hler metric on $\M_C$, and the $\bf \Lambda$ is a rotation matrix in symplectic space. The bosonic part of the action is:
\bea
    S_5  &=& \int\limits_5 {\left[ {R\star \mathbf{1} - \frac{1}{2}d\sigma \wedge\star d\sigma  - G_{i\bar j} dz^i \wedge\star dz^{\bar j} } \right.}  + e^\sigma   \left\langle {d\Xi } \right|\mathop{\bf \Lambda} \limits_ \wedge  \left| {\star d\Xi } \right\rangle\nonumber\\
    & &\left. {\quad\quad\quad\quad\quad\quad\quad\quad\quad\quad\quad\quad\quad - \frac{1}{2} e^{2\sigma } \left[ {da + \left\langle {\Xi } \mathrel{\left | {\vphantom {\Xi  {d\Xi }}} \right. \kern-\nulldelimiterspace} {{d\Xi }}    \right\rangle} \right] \wedge \star\left[ {da + \left\langle {\Xi } \mathrel{\left | {\vphantom {\Xi  {d\Xi }}} \right. \kern-\nulldelimiterspace} {{d\Xi }}    \right\rangle} \right] } \right],\label{action}
\eea
where $\star$ is the $D=5$ Hodge duality operator. The complete action, with fermionic fields, is symmetric under a set of SUSY transformations, as may be reviewed in \cite{Emam:2013mq}; where the following solution, representing one-dimensional branes coupled to the full hypermultiplet sector was found:

\bea
    ds^2  &=&  - dt^2 + dx^2  + f\left(r\right) \left( {dr^2  + r^2 d\theta ^2  + r^2 \sin ^2 \theta d\varphi ^2 } \right)     \label{1brane metric}\\
    \sigma \left( r \right) &=& \ln f\left(r\right), \quad\quad\quad
    a = a_\infty   \pm \frac{{7q}}{{2\sqrt 7 \left( {r  + q} \right)}},\quad {\rm where} \quad a_\infty \in \mathbb{R}\\
    dz^i  &=& - q f^i\frac{{dr}}{{r^2 }}\,\,\,\,\,\,{\rm such}\,\,{\rm that}\,\,\,\,\,\,df^i  - q\Gamma _{jk}^i f^j f^k\frac{dr}{r^2} =0\,\,\,\,\,\,{\rm and}\,\,\,\,\,\,G_{i\bar j} f^i f^{\bar j}  = \frac{{r^2 }}{{\left( {r  + q} \right)^2 }}    \\
    \left| \Xi  \right\rangle  &=&   \sqrt {\frac{r}{{7\left( {r + q} \right)}}} \mathfrak{Re}\left[ {\left( {1 \pm i\sqrt 7 } \right)\left| V \right\rangle } \right] + 2\sqrt {\frac{{r + q}}{{7r}}} \mathfrak{Re}\left[ {f^i \left| {U_i } \right\rangle } \right]\\
    \left| {d\Xi } \right\rangle  &=& \pm q\mathfrak{Re} \left[ {\frac{{\left( { \pm \sqrt 7  - i} \right)}}{{\sqrt {r\left( {r + q} \right)^3 } }}\left| V \right\rangle  + \frac{{2i}}{{\sqrt {r^3 \left( {r + q} \right)}  }}f^i \left| {U_i } \right\rangle } \right]dr,
\eea
where $x$ is the direction of the brane, $r$ is the radial direction in the bulk orthogonal to the brane, the functions $f^i$ are symplectic scalars, and the charge $q$ is an arbitrary real number with dimensions of length. The brane's metric (\ref{1brane metric}) is characterized by the warp function $f\left(r\right)=\left( {m + \frac{q}{{r }}} \right)$. The real integration constant $m$ is the value of $f$ at $r\rightarrow \infty$ which we will set to unity to ensure the asymptotic flatness of the metric.

\section{The Killing symmetries} \label{Killing}

A Killing vector field (or just Killing field) is a vector field on a Riemannian or pseudo-Riemannian manifold that preserves the metric. It defines isometries of the metric, which from the point of view of physics, leads to conserved quantities on the geodesics of the spacetime in question. Killing's equation, defined in terms of Lie derivatives, is
\be\label{Killing Equation}
    L_\xi g_{\mu\nu} = \nabla_\mu \xi_\nu + \nabla_\nu \xi_\mu = \partial_\mu \xi_\nu + \partial_\nu \xi_\mu + 2 \Gamma^\rho_{\mu\nu}\xi_\rho= 0,
\ee
where $\nabla$ is a Levi-Civita connection, and the Christoffel symbols of (\ref{1brane metric}) are:
\bea
    \Gamma^r_{rr} &=& -\frac{q}{2 r (q+r)}, \quad\quad\quad\quad\,\,\,
    \Gamma^r_{\theta\theta} = -\frac{r (q+2 r)}{2 (q+r)}, \nonumber\\
    \Gamma^r_{\phi\phi} &=& -\frac{r  (q+2 r)}{2 (q+r)} \sin ^2\theta, \quad\quad  \Gamma^\theta_{\phi\phi} = -\sin \theta  \cos \theta  \nonumber\\
    \Gamma^\theta_{r\theta} &=& \Gamma^\theta_{\theta r} = \frac{q+2 r}{2 q r+2 r^2}, \quad
    \Gamma^\phi_{r\phi} = \Gamma^\phi_{\phi r} = \frac{q+2 r}{2 q r+2 r^2}, \quad
    \Gamma^\phi_{\theta\phi} = \Gamma^\phi_{\phi\theta} = \cot \theta .
\eea

Hence, equation (\ref{Killing Equation}) leads to the set:
\bea\label{KILLING SPECIAL}
    \partial_t \xi_{t}&=&0, \quad\quad    \partial_x \xi_{x}=0, \quad\quad    \partial_t \xi_{x}+\partial_x\xi_{t}=0, \label{11}\\
    \partial_t \xi_{r}+\partial_r \xi_{t}&=&0, \quad\quad    \partial_x\xi_{r}+\partial_r\xi_{x}=0, \quad\quad    \partial_t \xi_{\theta}+\partial_\theta\xi_{t}=0, \label{22}\\
    \partial_x\xi_{\theta}+\partial_\theta\xi_{x}&=&0, \quad\quad    \partial_t \xi_{\phi}+\partial_\phi\xi_{t}=0, \quad\quad \partial_x\xi_{\phi}+\partial_\phi\xi_{x}=0, \label{33}\\
    \partial_r\xi_{r}+ \frac{q}{{2r\left( {q + r} \right)}}\xi_{r}&=&0, \quad\quad
    \partial_r\xi_{\theta}+\partial_\theta\xi_{r}-\frac{{q + 2r}}{{r\left( {q + r} \right)}} \xi_{\theta}=0, \label{44}\\
    \partial_r\xi_{\phi}+\partial_\phi\xi_{r}-\frac{{q + 2r}}{{r\left( {q + r} \right)}}\xi_{\phi}&=&0,\quad\quad
    \partial_\theta \xi_{\theta}+\frac{{r\left( {q + 2r} \right)}}{{2\left( {q + r} \right)}}\xi_{r}=0, \label{55}\\
    \partial_\theta\xi_{\phi}+\partial_\phi\xi_{\theta}-2\cot{\theta}\xi_{\phi} &=&0, \quad\quad
    \partial_\phi \xi_{\phi}+ \frac{{r\left( {q + 2r} \right)}}{{2\left( {q + r} \right)}}\sin^2\theta\xi_r +\sin{\theta}\cos{\theta}\xi_{\theta}=0.\label{66}
\eea

Solving the first equation of (\ref{44}) we get
\be\label{EQN1}
    \xi_r = \sqrt {\frac{{q + r}}{r}} C_1 \left( {t,x,\theta ,\phi } \right),
\ee
where the $C_i$ are arbitrary functions in their arguments. From the first and second equations of (\ref{11})
\be
    \xi _t  = C_2 \left( {x,r,\theta ,\phi } \right),\,\,\,\,\,\,\,\,\,\,\xi _x  = C_3 \left( {t,r,\theta ,\phi } \right).
\ee

Using (\ref{EQN1}) in the second equation of (\ref{44}) and the first of (\ref{55}) we find
\be
    \xi _\theta   = \frac{1}{{r\left( {q + r} \right)}}C_4 \left( {t,x,r,\phi } \right),\,\,\,\,\,\,\,\,\xi _\phi   = \frac{1}{{r\left( {q + r} \right)}}C_5 \left( {t,x,r,\theta } \right).
\ee

Using these results in the third equation of (\ref{22}) and the second of (\ref{33}) leads to:
\bea
    \xi _t  &=& A,\,\,\,\,\,\,\xi _x  = B,\,\,\,\,\,\,\xi _r  = 0 \nonumber\\
    \xi _\theta   &=& \frac{1}{{r\left( {q + r} \right)}}C_4 \left( {x,r,\phi } \right),\,\,\,\,\,\,\,\,\xi _\phi   = \frac{1}{{r\left( {q + r} \right)}}C_5 \left( {x,r,\theta } \right),
\eea
which in the equations of (\ref{66}) lead to
\bea
    C_4 \left( {x,r,\phi } \right) &=&  - D\sin \phi  + E\cos \phi  \nonumber\\
    C_5 \left( {x,r,\theta } \right) &=&  - \sin \theta \cos \theta \left( {D\cos \phi  + E\sin \phi } \right) + F\sin ^2 \theta,
\eea
where $A$, $B$, $D$, $E$, and $F$ are constants of integration. The general solution is then
\bea
    \xi  &=& - A\partial _t  + B\partial _x  + \frac{1}{{\left( {q + r} \right)^2 }}\left( { - D\sin \phi  + E\cos \phi } \right)\partial _\theta   \nonumber\\
    &+& \frac{1}{{\left( {q + r} \right)^2 \sin ^2 \theta }}\left[ { - \sin \theta \cos \theta \left( {D\cos \phi  + E\sin \phi } \right) + F\sin ^2 \theta } \right]\partial _\phi.
\eea

This gives the following five Killing vectors of the one-brane solution ($\forall \,\,r \ne -|q|$)
\bea
    \xi^{\left(1\right)} &=& \partial_t\\
    \xi^{\left(2\right)} &=& \partial_x\\
    \xi^{\left(3\right)} &=& \frac{1}{{\left( {q + r} \right)^2 }} \partial_\phi.\\
    \xi^{\left(4\right)} &=& \frac{{ - 1}}{{\left( {q + r} \right)^2 }}\left[ {\sin \phi \partial _\theta   + \cot \theta \cot \phi \partial _\phi  } \right].\\
    \xi^{\left(5\right)} &=& \frac{1}{{\left( {q + r} \right)^2 }}\left[ {\cos \phi \partial _\theta   - \cot \theta \sin \phi \partial _\phi  } \right].
\eea

Using $\xi_\mu U^\mu=$constant, where $U$ is the 5-velocity vector $\left(\dot t, \dot x, \dot r, \dot \theta, \dot \phi\right)$, and an over-dot is a derivative with respect to an affine parameter $\lambda$, leads to
\bea
    \dot t = {\rm constant}, \quad\quad \dot x &=& {\rm constant}, \quad\quad \left( {1 + \frac{q}{r}} \right)r^2\sin^2\theta \dot \phi = {\rm constant}\label{Con1}\\
    r^2 \left( {\cos \phi  - \sin \phi } \right)\dot \theta  &=& {\rm constant}\label{Con2}\\
    - r^2 \left( {\cos \phi  + \frac{{\cos ^2 \phi }}{{\sin \phi }}} \right) &=& {\rm constant}.\label{Con3}
\eea

We identify the conserved quantities of (\ref{Con1}) as related to the conservation of energy, linear momentum along the one-brane, and the azimuthal angular momentum about the brane's axis respectively, while (\ref{Con2}) and (\ref{Con3}) are the other rotations in the three-dimensional bulk space.

\section{The geodesic equations}\label{}

In the metric (\ref{1brane metric}), the constant $q$ is the charge coupling the brane to the hypermultiplets. It may, in principle, acquire positive or negative values and we will explore both cases. For the positive case, the warp function is positive over its entire domain $\left\{ {r|r \ge 0} \right\}$, while the negative case introduces a spherical barrier at $r = R_b = \left| q \right|$. The metric inside this radius switches signature from $\left(-,+,+,+,+\right)$ to $\left(-,+,-,-,-\right)$ which signals a causal disconnect between the two regions. The Ricci scalar of this spacetime is
\be
    R= \frac{3 q^2}{2 r (q+r)^3}.
\ee

Clearly a positively charged one brane has a single naked singularity at $r=0$, while the negatively charged one has \emph{two} singularities, one on the brane at $r=0$, and one on a spherical shell of radius $r = -|q|$. For the negative solution, we then expect that the geodesics will be discontinued at this boundary. We will also see that it is a repulsive singularity. The geodesic equation
\be
     \frac{{d^2 x^\sigma  }}{{d\lambda ^2 }} + \Gamma _{\mu \nu }^\sigma  \frac{{dx^\mu  }}{{d\lambda }}\frac{{dx^\nu  }}{{d\lambda }} = 0, \quad\quad \mu ,\nu ,\sigma  = 0, \ldots ,4
\ee
leads to
\bea\label{GeoEqn1}
    \ddot r &=& \frac{{q\dot r^2 }}{{2r\left( {q + r} \right)}} + \frac{{r\left( {q + 2r} \right)}}{{2\left( {q + r} \right)}}\left( {\dot \theta ^2  + \dot \varphi ^2 \sin ^2 \theta } \right) \nonumber\\
    \ddot \theta  &=& \sin \theta \cos \theta \dot \varphi ^2  - \frac{{\dot r\dot \theta \left( {q + 2r} \right)}}{{r\left( {q + r} \right)}} \nonumber\\
    \ddot \varphi  &=&  - \left[ {\frac{{\dot r\left( {q + 2r} \right)}}{{r\left( {q + r} \right)}} + 2\cot \theta \dot \theta } \right]\dot \varphi.
\eea

Equations (\ref{GeoEqn1}) can be simplified by considering the Killing symmetries found in the previous section. We can use the third equation of (\ref{Con1}) to simplify (\ref{GeoEqn1}). In fact, since angular momentum conservation forces the geodesics to remain in a specific plane, we can choose $\theta=$constant to fix that plane. The second equation of (\ref{GeoEqn1}) then leads to $\sin \theta \cos \theta = 0$, which implies either $\theta=0^\circ$ or $\theta=90^\circ$, but the last equation diverges for the first choice, leading us to conclude that $\theta=90^\circ$; the equatorial plane with respect to the brane's axis. Hence the third conserved quantity in (\ref{Con1}) becomes
\be\label{AngMom}
    \left(1+\frac{q}{r}  \right)r^2 \dot \varphi  = l,
\ee
where $l$ is understood to be the conserved angular momentum parameter. Rewriting (\ref{AngMom}) to solve for $\dot \varphi$ gives:
\be\label{AngMom2}
    \dot \varphi  = \frac{l}{{r^2  + qr}}.
\ee


If $q=0$, $\dot \varphi$ is inversely proportional to $r^2$ as would be expected for a flat spacetime. For the case $q>0$, $\dot \varphi$ becomes smaller as $r$ gets larger; i.e. the effect of $q$ becomes more significant at large distances from the brane. While for $q<0$, the denominator gets smaller as $r$ increases, and $\dot \varphi$ gets larger. One must be careful however, since as the denominator approaches zero and subsequently becomes negative, this causes $\dot \varphi$ to become undefined at the two singularities of the brane and then changing sign. Since the second equation of (\ref{GeoEqn1}) now vanishes identically, and the third is just the derivative of (\ref{AngMom}), we are only left with
\be\label{rEqn}
    \ddot r = \frac{{q\dot r^2 }}{{2r\left( {q + r} \right)}} + \frac{{\left( {q + 2r} \right)l^2 }}{{2r\left( {q + r} \right)^3 }}.
\ee

\section{The effective potential}\label{}

The Newtonian effective potential method is particularly useful in understanding the possible geodesics of a particular spacetime. To see this, we observe that the first and second of the conserved quantities in (\ref{Con1}) can be used in the 5-velocity normalization condition as follows:
\bea
    U^\mu  U_\mu   &=&  - \dot t^2  + \dot x^2  + f\left( r \right)\dot r^2  + f\left( r \right)r^2 \dot \theta ^2  + f\left( r \right)r^2 \sin ^2 \theta \dot \varphi ^2  = \varepsilon  \\
    &=&  - E^2  + p^2  + f\left( r \right)\dot r^2  + f\left( r \right)r^2 \dot \varphi ^2  = \varepsilon,\label{VelocityNorm}
\eea
where $E$ and $p$ are the energy and linear momentum related constants, and we have defined $\varepsilon$ as $0$, $-1$, or $+1$ for null, timelike, and spacelike geodesics respectively. Using (\ref{AngMom}), as well as collecting the constants together $\E=E^2-p^2+\varepsilon$, gives
\be
    \E = f\left( r \right)\dot r^2  + \frac{{l^2 }}{{f\left( r \right)r^2 }},
    = \left( {q + r} \right)\frac{{\dot r^2 }}{r} + \frac{{l^2 }}{{r\left( {q + r} \right)}}.\label{EnCons}
\ee

In the manner familiar from similar calculations, we recognize the second term of (\ref{EnCons}) as the Newtonian effective potential:
\be\label{Eff}
    V_{eff}\left(r\right) = \frac{{l^2 }}{{f\left( r \right)r^2 }} = \frac{{l^2 }}{{r\left( {q + r} \right)}}.
\ee

\begin{figure}[!ht]
  \begin{subfigure}[b]{.5\linewidth}
    \centering
    \includegraphics[scale=0.6]{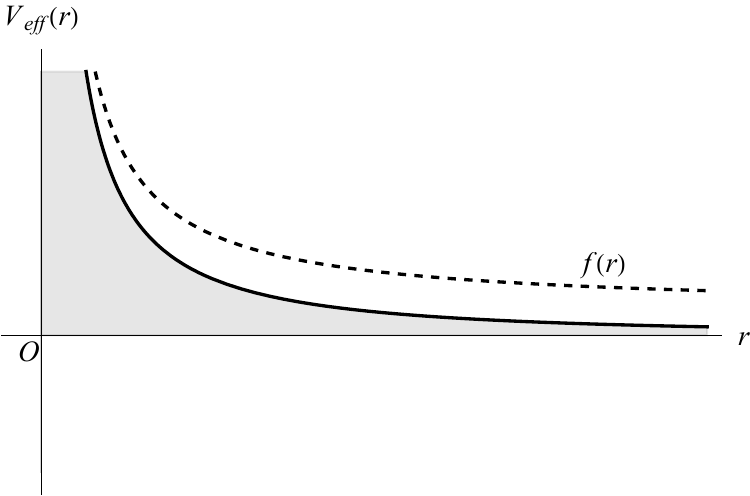}
    \caption{$q>0$; $q=+1$, $l=+1$, $r:\left( {0, 1.3} \right)$.}
    \label{Veffective1}
  \end{subfigure}%
  \begin{subfigure}[b]{.5\linewidth}
    \centering
    \includegraphics[scale=0.6]{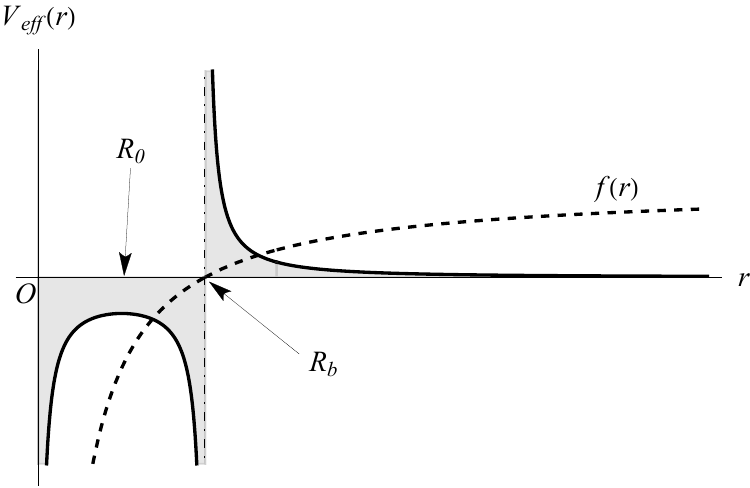}
    \subcaption{$q<0$; $q=-1$, $l=+1$, $r:\left( {0, 4} \right)$.}
    \label{Veffective2}
  \end{subfigure}
  \caption{The effective potentials and the warp function (not to scale). The `forbidden' regions are shaded. }
  \label{fig:1}
\end{figure}

Clearly then, the quantity
\be
    \dot r^2  = \frac{\E-V_{eff}}{f}
\ee
cannot be allowed to become negative, which puts a restriction on the behavior of the geodesics. The form of $V_{eff}$ is shown in Fig. \ref{fig:1} for both the cases of positive and negative $q$. Inspection of $V_{eff}$ indicate that for $q>0$ there are no closed orbital geodesics, all are either open or radial, while for the case $q<0$ there are possible closed orbital geodesics for the region $r<R_b$ with exactly one circular geodesic at $r=R_0$. We will discuss the stability of this last one in the next section.

It is worthwhile to note here that the geodesic \emph{structure} is the \emph{same} for all types geodesics; null, timelike, or spacelike. This can be seen by choosing the effective potential to depend on $\varepsilon$ as follows:
\be
    V_{eff}\left(r, \varepsilon\right) = \frac{{l^2 }}{{f\left( r \right)r^2 }} - \varepsilon.
\ee

This implies that the effective potential will only `shift' vertically by a factor of 1 for the timelike case, but will have the same structure as (\ref{Eff}) and as such the exact same geodesics for the various boundary conditions. This is similar to some prior work as in \cite{Teo:2003,Frolov:1998,Bardeen:1972}.

\section{Stability of circular orbits}\label{}

The value of the radius $R_0$ of the circular orbit for the case of negative $q$ may be found in the familiar manner of setting
\be
    V'_{eff}  = \frac{{dV_{eff} }}{{dr}} =  - \frac{{l^2 \left( {q + 2r} \right)}}{{r^2 \left( {q + r} \right)^2 }}
\ee
to zero and solving for $r=R_0$, or by setting $\ddot r = \dot r = 0$ in (\ref{rEqn}). Both of which give $R_0  = - {q \mathord{\left/
 {\vphantom {q 2}} \right. \kern-\nulldelimiterspace} 2} = {{\left| q \right|} \mathord{\left/
{\vphantom {{\left| q \right|} 2}} \right. \kern-\nulldelimiterspace} 2}$. The stability of this orbit is investigated by perturbatively disturbing the circular orbit. This is done by setting $r=R_0+\epsilon\left( \lambda  \right)$, where $\epsilon<<R_0$ is a small perturbation, in equation (\ref{rEqn}). Expanding and ignoring terms of $O\left(\epsilon^2\right)$ and $O\left(\dot\epsilon^2\right)$ leads to, after a bit of algebra:
\be
    \ddot \varepsilon  = \frac{{l^2 }}{{2R_0^2 \left( {q + R_0 } \right)^4 }}\left[ {R_0 \left( {q + R_0 } \right)\left( {q + 2R_0 } \right) - \left( {q^2  + 4qR_0  + 6R_0^2 } \right)\varepsilon } \right],
\ee
which ends up being
\be
    \ddot \varepsilon  + \left( {\frac{{l^2 }}{{R_0^4 }}} \right)\varepsilon  = \ddot \varepsilon  + \left( {\frac{{16l^2 }}{{q^4 }}} \right)\varepsilon  = 0.
\ee

This of course has the general solution
\be
    \epsilon \left( \lambda  \right) = A\sin \left( {\omega \lambda } \right) + B\cos \left( {\omega \lambda } \right),
\ee
where $A$ and $B$ are arbitrary constants and $\omega  = {{l } \mathord{\left/ {\vphantom {{l^2 } {R_0^2 }}} \right. \kern-\nulldelimiterspace} {R_0^2 }} = {{4l } \mathord{\left/ {\vphantom {{4l } {q^2 }}} \right. \kern-\nulldelimiterspace} {q^2 }}$ is the angular frequency of this oscillatory solution. Hence the circular orbits at $R_0$ are stable, despite the fact that the potential looks `upside down' at this point. This behavior is due to the change of signature of the metric inside the region $r<R_b$.

\section{The radial geodesics}\label{}

We now study the radial geodesics. The energy conservation relation (\ref{EnCons}) reduces to
\be
    \dot r^2  = \frac{\E}{f},\label{energy2}
\ee

Equation (\ref{energy2}) leads to
\be
    \lambda\left( r \right) = \int {dr\sqrt {\frac{f}{\E}} }
\ee
the solution of which is
\be\label{RadGeo}
    \lambda \left( r \right) = r\sqrt {\frac{f}{\E}}  + \frac{q}{{\sqrt \E }}\tanh ^{ - 1} \left( {\frac{1}{{\sqrt f }}} \right).
\ee

\begin{figure}[H]
  \begin{subfigure}[b]{.5\linewidth}
    \centering
    \includegraphics[scale=0.4,left]{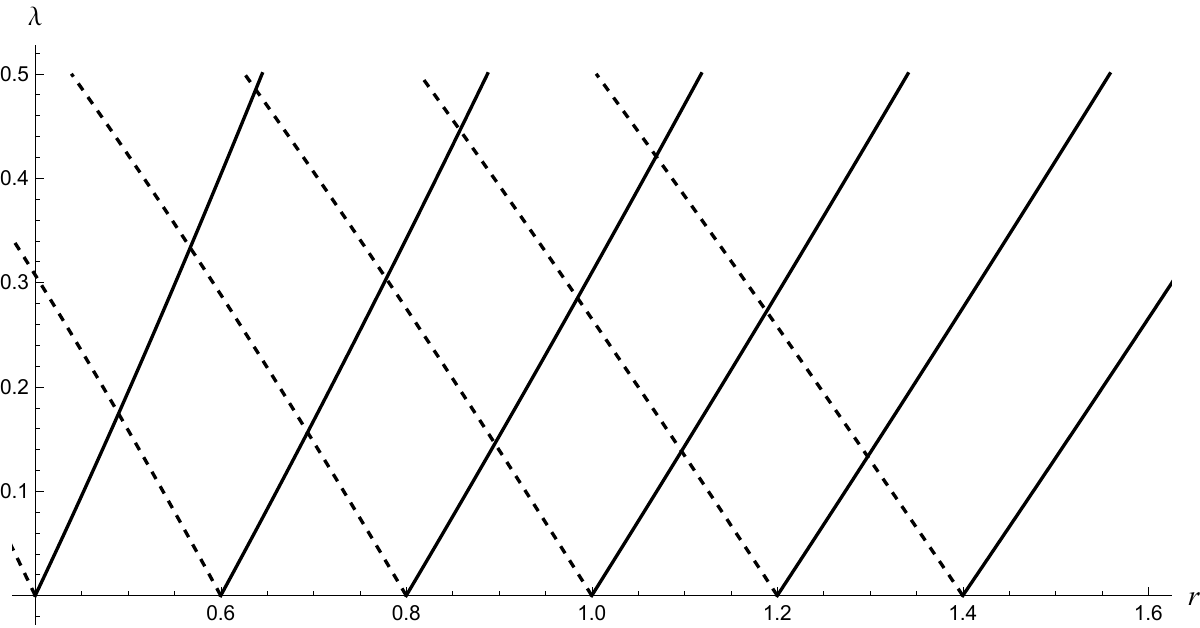}
    \caption{Radial geodesics for $q=+1$.}
    \label{fig:4a}
  \end{subfigure}
  \begin{subfigure}[b]{.5\linewidth}
    \centering
    \includegraphics[scale=0.4,right]{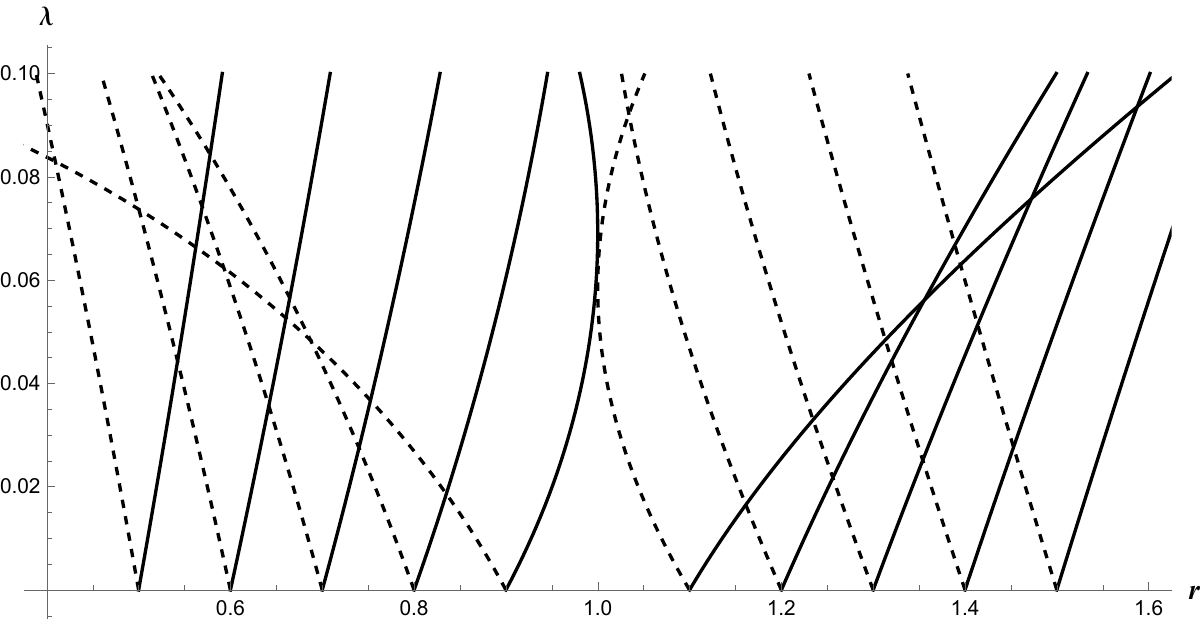}
    \subcaption{Radial geodesics for $q=-1$.}
    \label{fig:4b}
  \end{subfigure}
  \caption{Comparison of the radial geodesics of positive and negative $q$. The disconnect at $r=R_b=1$ is clear for the negative case.}
  \label{fig:4}
\end{figure}

\begin{figure}[H]
  \begin{subfigure}[b]{.5\linewidth}
    \centering
    \includegraphics[scale=0.5,center]{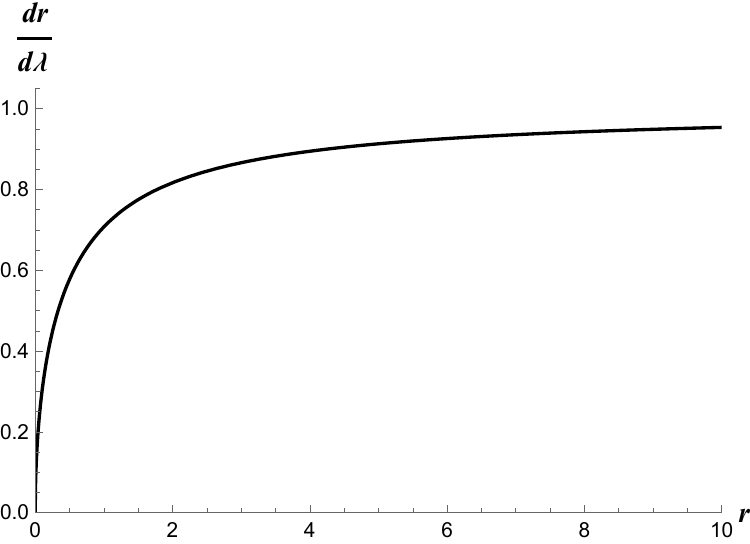}
    \caption{$q=+1$.}
    \label{fig:5a}
  \end{subfigure}
  \begin{subfigure}[b]{.5\linewidth}
    \centering
    \includegraphics[scale=0.5,center]{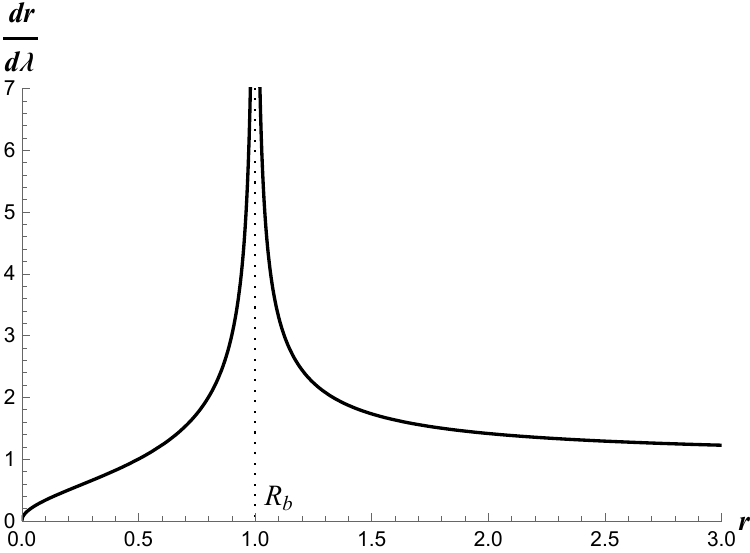}
    \subcaption{$q=-1$.}
    \label{fig:5b}
  \end{subfigure}
  \caption{$\dot r$ vs r.}
  \label{fig:5}
\end{figure}

For the positive $q$ solution this expression is well behaved for all $\E>0$, since $f$ is always positive. For the negative $q$ case (\ref{RadGeo}) is well behaved for $r>R_b$ \emph{and} $\E>0$, while for $r<R_b$, where $f$ becomes negative, the first term is well behaved if $\E<0$, and the inverse $\tanh$ function in the second term becomes
\be
    \tanh ^{ - 1} \left( {\frac{1}{{\sqrt f }}} \right) = \tanh ^{ - 1} \left( {\frac{{ - i}}{{\sqrt {\left| f \right|} }}} \right) =  - i\tan ^{ - 1} \left( {\frac{1}{{\sqrt {\left| f \right|} }}} \right),
\ee
which requires $\E<0$ as well. The geodesics in the two regions of the negative $q$ case are then clearly disconnected. This is demonstrated in Fig. \ref{fig:4} where we have plotted the radial geodesics for both positive and negative $q$ for comparison. It is interesting to note that the shell singularity in the $q<0$ case seems to `repel' the geodesics directed towards it. This behavior will also manifest itself in the orbital geodesics as we will see in the next section. For further comparison, we also plot a sample of the `velocity' equation (\ref{energy2}) for both cases in Fig. \ref{fig:5}. The rate of change of $r$ {w.r.t} $\lambda$ asymptotes to 1 at radial infinity for both cases as expected. In the negative $q$ case the disconnect at $r=R_b$ shows clearly.

\section{The orbital (bound and unbound) geodesics}\label{}

\begin{figure}[H]
  \begin{subfigure}[b]{.5\linewidth}
    \centering
    \includegraphics[scale=0.4]{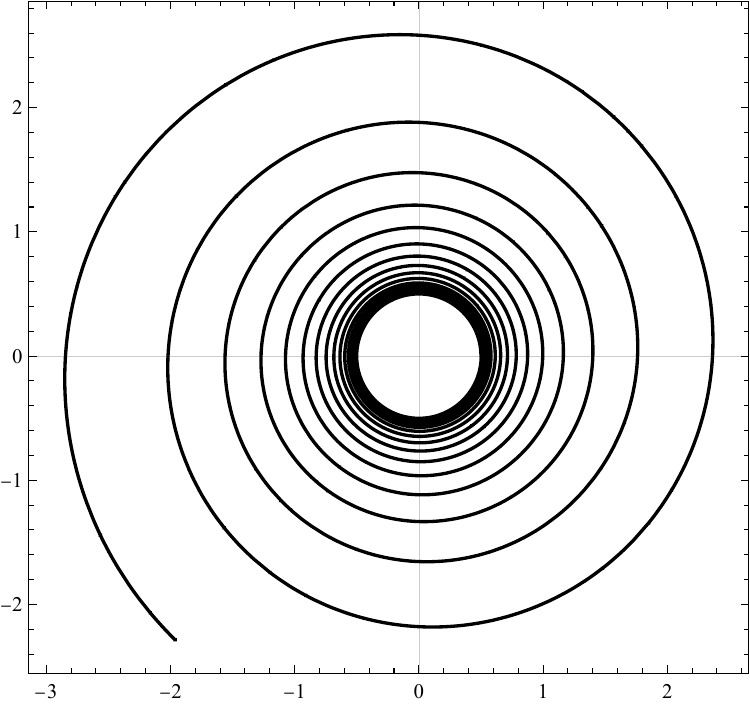}
    \caption{Orbital with vanishing initial $\dot r$.}
    \label{1}
  \end{subfigure}
  \begin{subfigure}[b]{.5\linewidth}
    \centering
    \includegraphics[scale=0.6]{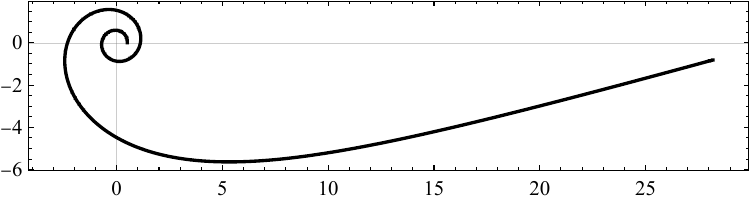}
    \subcaption{Orbital with positive initial $\dot r = + 2.8$ (outgoing).}
    \label{2}
  \end{subfigure}
  \begin{subfigure}[b]{.5\linewidth}
    \centering
    \includegraphics[scale=0.4]{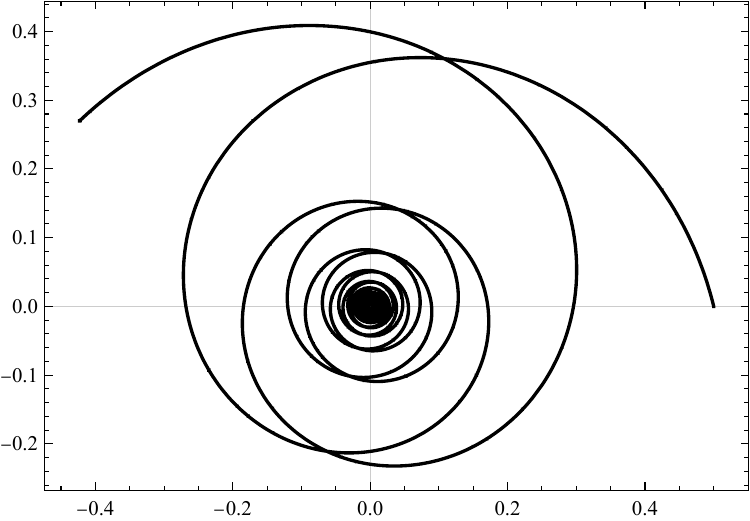}
    \caption{Orbital with negative initial $\dot r = - 7$ (ingoing). The geodesics `bounce' back at the potential boundary as expected.}
    \label{3}
  \end{subfigure}
  \caption{Results of numerically solving equations (\ref{GeoEqn1}) for $q=+1$. For all three plots the initial values are $r=0.5$, $\varphi=0$, $\dot \varphi=1$, and $t:\left( {0,5} \right)$}
  \label{fig:2}
\end{figure}

We now attempt to numerically solve equations (\ref{GeoEqn1}) for $\theta=90^\circ$. We begin with the trivial case of positive $q$, outlined in Fig. \ref{fig:2} for various initial `velocities.' As expected, all of the orbital geodesics are unbound for this case.

For the, more interesting, case of negative $q$ we treat the problem as follows: First we set $l=-1$ for a counterclockwise plot, then $\dot r=0$ in equation (\ref{EnCons}) to find the energy $\E_0$ of the circular orbit at $r=R_0$:
\be
    \E_0  = \frac{{l^2 }}{{f\left( {R_0 } \right)R_0^2 }}.
\ee

By choosing appropriate values for the initial radial distance $r_i$ we also calculate the initial radial velocity for specific values of $\E$ by
\be
    \dot r_i  =  \pm \sqrt {\frac{{\E  - V_{eff} \left( {r_i ,\ell } \right)}}{{f\left( {r_i } \right)}}}.
\ee

Using these last two equations we can now classify the solutions by their initial radii, energies, and velocities as shown in Table \ref{tab:my-table}.

\begin{figure}[hp]
  \begin{subfigure}[b]{.5\linewidth}
    \centering
    \includegraphics[scale=0.5]{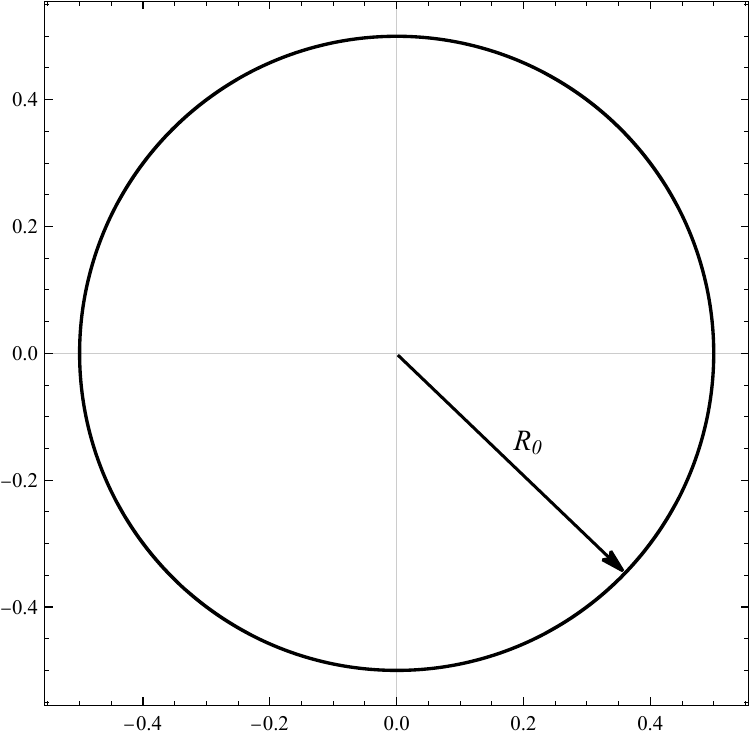}
    \caption{The circular orbit, $t:\left( {0,1} \right)$.}
    \label{1}
  \end{subfigure}
  \begin{subfigure}[b]{.5\linewidth}
    \centering
    \includegraphics[scale=0.5]{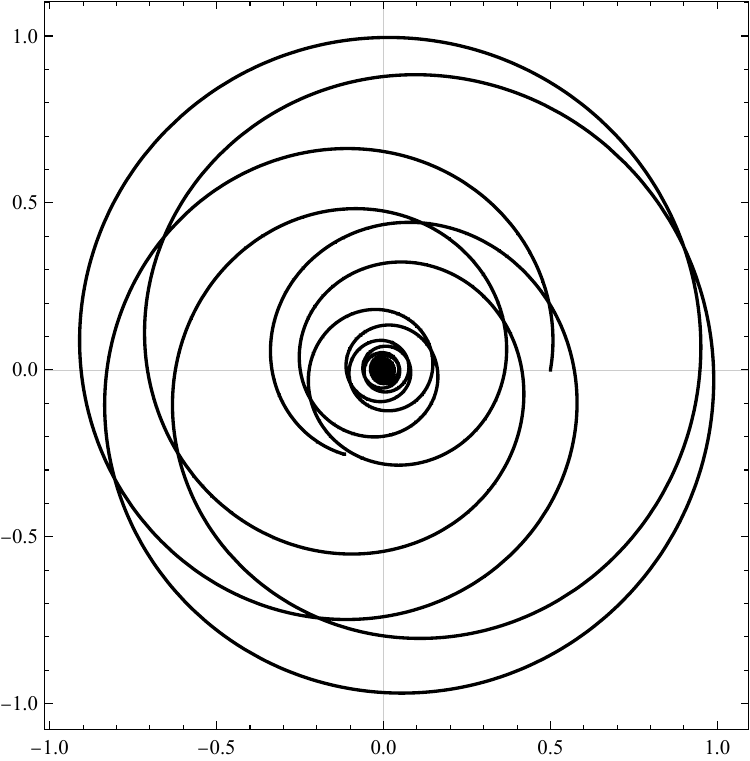}
    \subcaption{An inner orbit with low energy, $t:\left( {0,0.2} \right)$.}
    \label{2}
  \end{subfigure}
  \begin{subfigure}[b]{.5\linewidth}
    \centering
    \includegraphics[scale=0.5]{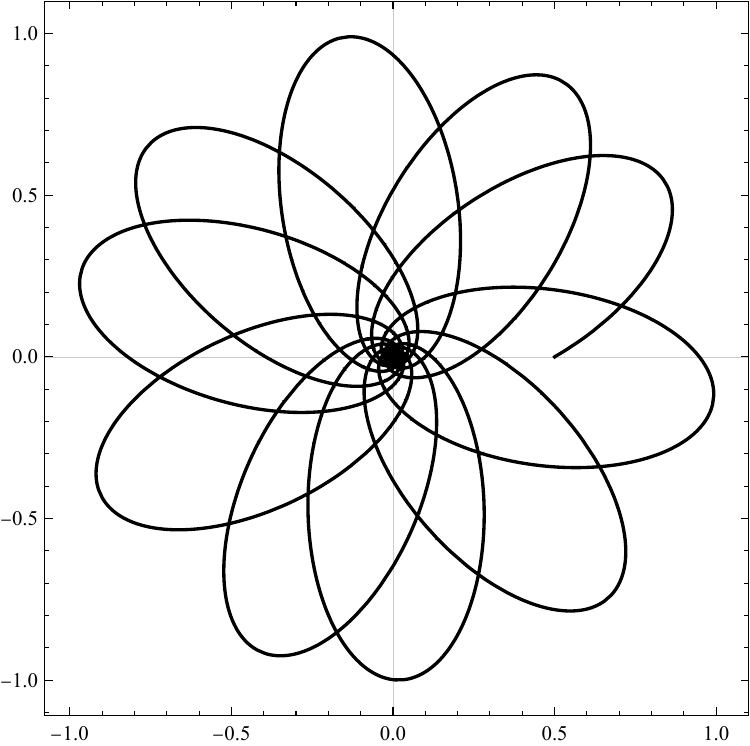}
    \caption{An inner orbit with high energy, $t:\left( {0,0.15} \right)$.}
    \label{3}
  \end{subfigure}
  \begin{subfigure}[b]{.5\linewidth}
    \centering
    \includegraphics[scale=0.4]{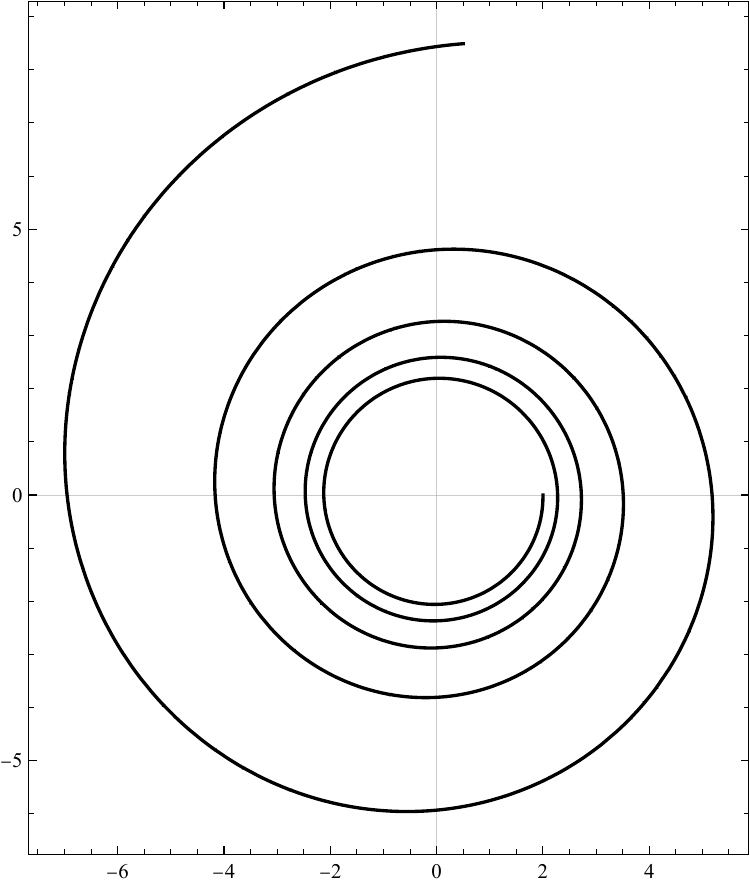}
    \caption{An outer region geodesic: $\dot r_i > 0$, $t:\left( {0,6} \right)$. }
    \label{4}
  \end{subfigure}
  \begin{subfigure}[b]{.5\linewidth}
    \centering
    \includegraphics[scale=0.5]{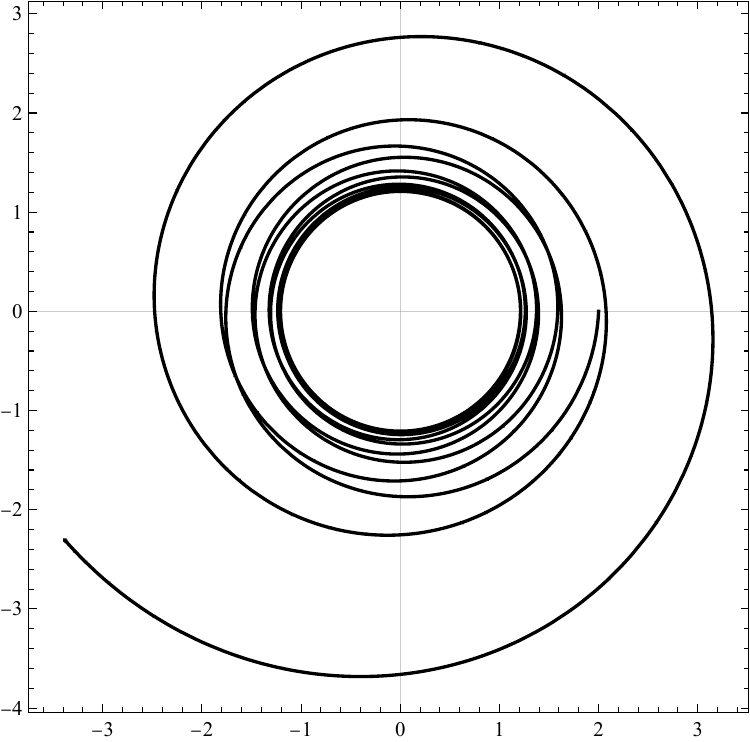}
    \caption{An outer region geodesic: $\dot r_i < 0$, $t:\left( {0,1.5} \right)$.}
    \label{5}
  \end{subfigure}
  \begin{subfigure}[b]{.5\linewidth}
    \centering
    \includegraphics[scale=0.5]{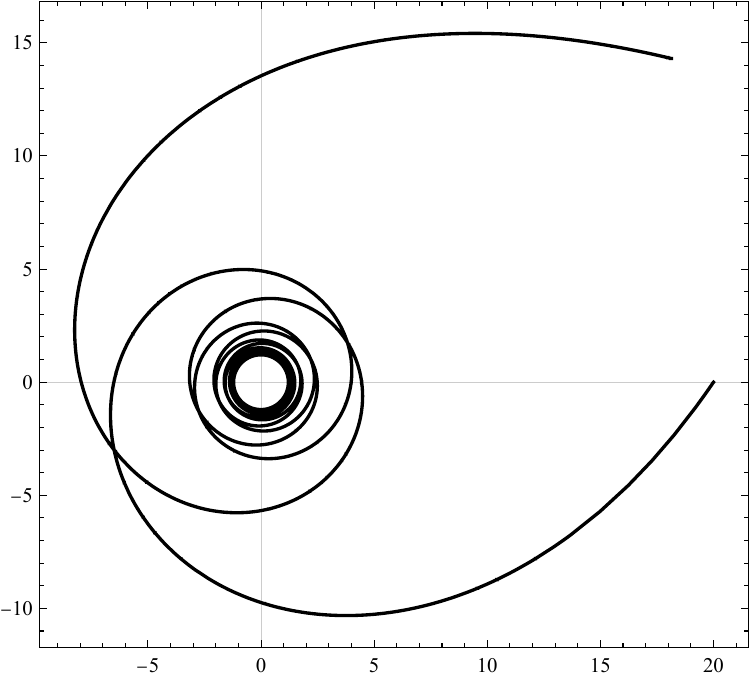}
    \caption{A slingshot geodesic: $\dot r_i <0$, $r_i \rightarrow \infty $, $t:\left( {0, 19} \right)$.}
    \label{6}
  \end{subfigure}
  \caption{Results of numerically solving equations (\ref{GeoEqn1}) for $q=-1$, $l=+1$ as classified in Table \ref{tab:my-table}.}
  \label{fig:3}
\end{figure}

\begin{figure}[!ht]
  \begin{subfigure}[b]{.5\linewidth}
    \centering
    \includegraphics[scale=0.4]{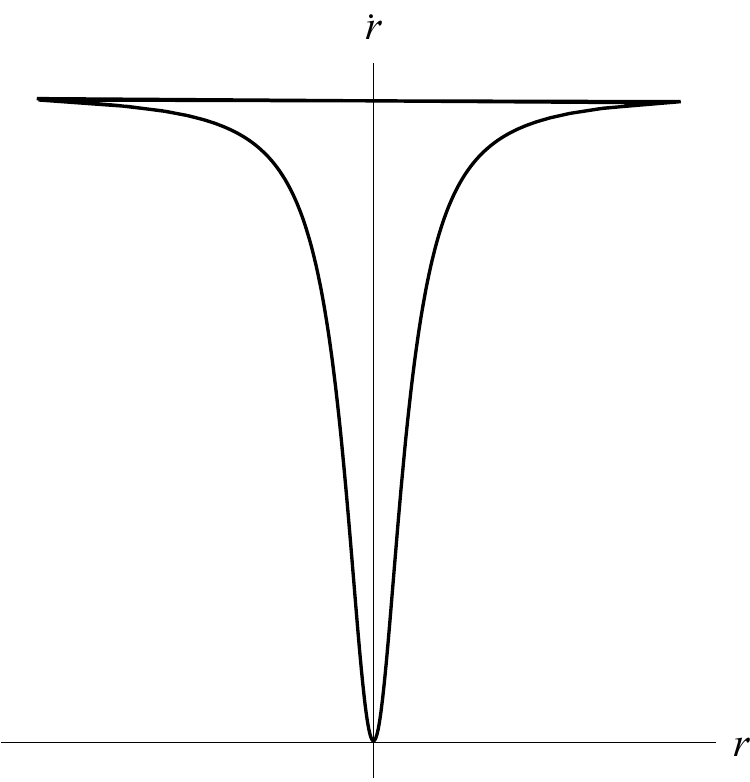}
    \caption{The phase diagram for figure (\ref{2})}
    \label{P1}
  \end{subfigure}%
  \begin{subfigure}[b]{.5\linewidth}
    \centering
    \includegraphics[scale=0.4]{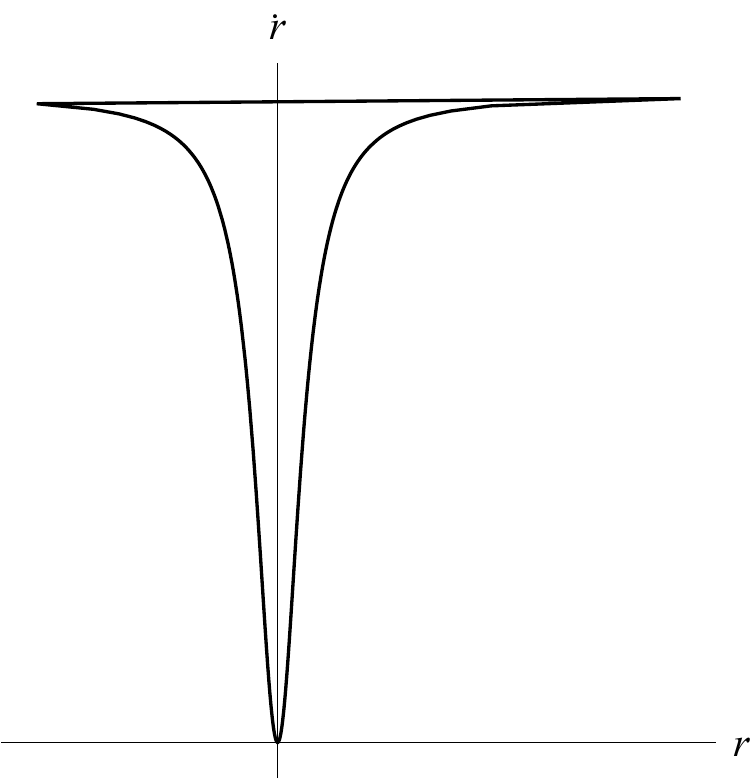}
    \subcaption{The phase diagram for figure (\ref{3})}
    \label{P2}
  \end{subfigure}
  \caption{Phase diagrams $\dot r$ vs $r$ for the inner orbits.}
  \label{fig:6}
\end{figure}

The results are shown in Fig. \ref{fig:3}. In Fig. \ref{fig:6} we also provide phase diagrams $\dot r$ vs $r$ to demonstrate that the bound inner region geodesics are closed.

\begin{table}[]
\centering
\begin{tabular}{|c|c|c|c|c|}
\hline
$q=-1$                        & Figure    & $r_i$                   & $\E$        & $\dot r_i$ \\ \hline
\multirow{3}{*}{Inner region} & \ref{1}   & \multirow{3}{*}{$R_0$}  & $\E_0$      & $0$        \\ \cline{2-2} \cline{4-5}
                              & \ref{2}   &                         & $100\E_0$   & $>0$       \\ \cline{2-2} \cline{4-5}
                              & \ref{3}   &                         & $10,000\E_0$ & $>0$       \\ \hline
\multirow{3}{*}{Outer region} & \ref{4}   & \multirow{2}{*}{$2R_b$} & $-\E_0/4$   & $>0$       \\ \cline{2-2} \cline{4-5}
                              & \ref{5}   &                         & $-\E_0$     & $<0$       \\ \cline{2-5}
                              & \ref{6}   & $20R_b$                 & $-\E_0$     & $<0$       \\ \hline
\end{tabular}
\caption{The $q=-1$ solutions classified by energy and initial conditions.}
\label{tab:my-table}
\end{table}

\section*{Conclusion}
This work is a continuation of the results presented in our previous paper \cite{Emam:2013mq}, whose primary objective was to apply the methods developed in \cite{Emam:2009xj} and construct $D=5$ hypermultiplet fields in a specific spacetime background simply by exploiting the symplectic symmetry of the theory and finding solutions that are based on symplectic invariants and vectors. In this paper, we continued this work by finding the fully calculating the geodesic structure of that spacetime. The results are characterized by the coupling constant $q$. For the case of positive $q$ the geodesics are smooth over all the bulk space $r>0$, while for negative $q$ a singular spherical barrier exists and acts to `repel' the geodesics. This is in contrast to similar solutions where the singularity acted as an attractor, such as \cite{Chandler:2015aha}.

\end{document}